\documentclass[11pt]{article}
\usepackage{amsmath,epsf,amssymb,latexsym,enumerate,cite,shadow,array,color}
\usepackage[ps,dvips,matrix,arrow,frame,import,curve,color]{xy}
\renewcommand{\theequation}{\thesection.\arabic{equation}}
\newcounter{saveeqn}
\newcommand{\add}{\addtocounter{equation}{1}}
\newcommand{\alpheqn}{\setcounter{saveeqn}{\value{equation}}%
\setcounter{equation}{0}%
\renewcommand{\theequation}{\mbox{\thesection.\arabic{saveeqn}{\alph{equation}}}}}
\newcommand{\reseteqn}{\setcounter{equation}{\value{saveeqn}}%
\renewcommand{\theequation}{\thesection.\arabic{equation}}}

\newif\iffigs\figstrue

\DeclareFontFamily{U}{rsf}{}
\DeclareFontShape{U}{rsf}{m}{n}{
  <5> <6> rsfs5 <7> <8> <9> rsfs7 <10-> rsfs10}{}
\DeclareMathAlphabet\Scr{U}{rsf}{m}{n}

\def\pplogo{\vbox{\kern-\headheight\kern -29pt
\halign{##&##\hfil\cr&{
\ppnumber}\cr\rule{0pt}{2.5ex}&\ppdate\cr}
}}
\makeatletter
\def\ps@firstpage{\ps@empty \def\@oddhead{\hss\pplogo}%
  \let\@evenhead\@oddhead 
}
\def\maketitle{\par
 \begingroup
 \def\thefootnote{\fnsymbol{footnote}}
 \def\@makefnmark{\hbox{$^{\@thefnmark}$\hss}}
 \if@twocolumn
 \twocolumn[\@maketitle]
 \else \newpage
 \global\@topnum\z@ \@maketitle \fi\thispagestyle{firstpage}\@thanks
 \endgroup
 \setcounter{footnote}{0}
 \let\maketitle\relax
 \let\@maketitle\relax
 \gdef\@thanks{}\gdef\@author{}\gdef\@title{}\let\thanks\relax}
\makeatother

\newcommand{\bea}{\begin{eqnarray}}
\newcommand{\eea}{\end{eqnarray}}
\newcommand{\be}{\begin{equation}}
\newcommand{\ee}{\end{equation}}

\begin{document}

\thispagestyle{empty}

\begin{titlepage}
\begin{flushright}
SU-ITP-2011/09\\
\end{flushright}


\vskip  2 cm

\vspace{24pt}

\begin{center}
{ \LARGE \textbf{     $E_{7(7)}$ Symmetry and Finiteness of ${\cal{N}}=8$ Supergravity
 }}

\vspace{28pt}

{\bf Renata Kallosh}

\

 \textsl{Department of Physics,
    Stanford University}\\ \textsl{Stanford, CA 94305-4060, USA}

\vspace{10pt}

\vspace{24pt}

\end{center}

\begin{abstract}

We study ${\cal{N}}=8$ supergravity deformed by the presence of  the candidate counterterms. We show that even though  they are  invariant under undeformed $E_{7(7)}$, all of the candidate counterterms violate the  deformed $E_{7(7)}$ current conservation. The same conclusion follows from the uniqueness of the Lorentz and $SU(8)$ covariant, $E_{7(7)}$ invariant unitarity constraint expressing the 56-dimensional  $E_{7(7)}$ doublet via 28 independent vectors, in agreement with the ${E_{7(7)}\over SU(8)}$ coset space geometry. Therefore  $E_{7(7)}$ duality predicts the all-loop UV finiteness of perturbative ${\cal{N}}=8$ supergravity.

\end{abstract}

\end{titlepage}

\newpage

\section{Introduction}

Recently we argued in \cite{Kallosh:2010kk}  that ${\cal{N}}=8$ supergravity \cite{Cremmer:1979up} is perturbatively UV finite. The proof was rather complicated, based on  a relation between the
real and chiral versions of the off-shell light-cone superspace. It  generalized  the perturbative supergraph  non-renormalization theorem for the superpotential.  It was necessary to compare properties of the light-cone superspace with that of the covariant superspace   and combine it all with the recent approach based on the helicity amplitude computations.

In this paper we will study this issue using a simpler set of  more familiar tools, which do not require knowledge of the off-shell light-cone superspace and helicity amplitudes. We will  analyze all candidate counterterms  \cite{Howe:1980th,Kallosh:1980fi} using the standard Lorentz covariant on shell superspace approach \cite{Brink:1979nt}. We will show that the power of the continuous global $E_{7(7)}(\mathbb{R})$  duality symmetry  and of the corresponding current conservation was underestimated. It is, in fact, strong enough to forbid all perturbative\footnote{The $E_{7(7)}(\mathbb{R})$ symmetry is expected to be a symmetry of perturbative  ${\cal{N}}=8$ supergravity.  It is broken to an arithmetic subgroup  $E_{7(7)}(\mathbb{Z})$ by non-perturbative effects. However, one can still use $E_{7(7)}(\mathbb{R})$  symmetry for investigation of UV divergences in the perturbation theory; see \cite{Bianchi:2009wj} for the recent discussion of this issue.} UV divergences which are consistent with the gauge symmetries of the theory.

The discovery of the 3-loop UV finiteness \cite{Bern:2007hh} of ${\cal{N}}=8$ supergravity  attracted attention to a possibility that  the hidden $E_{7(7)}$ symmetry of the theory may be relevant to the UV properties of the theory \cite{Brink:2008qc,Kallosh:2008ic,Bianchi:2008pu,Kallosh:2008ru}.   The full set of non-linear $E_{7(7)}$ invariant counterterms was constructed long time ago in \cite{Howe:1980th,Kallosh:1980fi}, based on ${\cal{N}}=8$ supergravity in the on shell superspace  \cite{Brink:1979nt}.
Recently the explanation of the 3-loop UV finiteness
due to unbroken $E_{7(7)}$ continuous symmetry was proposed in \cite{Brodel:2009hu}.  $E_{7(7)}$ constraints on linearized counterterms in ${\cal{N}}=8$ supergravity were studied in \cite{Beisert:2010jx}. A recent review of the candidate counterterms  from the amplitude viewpoint is available in \cite{Elvang:2010xn}.

Thus, it was  recognized that $E_{7(7)}$ symmetry may impose stringent constraints on the structure of the candidate counterterms. However, the standard lore was that as long as the candidate counterterms are $E_{7(7)}$ invariant, they should be allowed by the $E_{7(7)}$ symmetry of the theory. As we are going to show, this is not the case; adding $E_{7(7)}$ invariant counterterms to the theory may break the $E_{7(7)}$ symmetry of the theory deformed by the addition of these counterterms.

  The unusual properties of  continuous global duality symmetries make the standard Noether construction  not useful for the vector field part of the symmetry. To study duality symmetries one has to use the Noether-Gaillard-Zumino type construction \cite{Gaillard:1981rj,Aschieri:2008ns}, which provides the conserved current for such symmetries and guaranties that the equations of motion are  duality invariant.\footnote{The {\it bona fide} Noether current of the Lorentz non-covariant version of the theory developed in \cite{Bossard:2010dq} and the Hamiltonian approach in \cite{Bunster:2011aw}  may be used for an alternative analysis of the UV properties of  ${\cal{N}}=8$ supergravity. }
 In application to ${\cal{N}}=8$ supergravity, the current of the classical theory was constructed explicitly in \cite{Kallosh:2008ic} by embedding $E_{7(7)}$ into $Sp(2n,{\mathbb{R}})$. In this paper we derive a particularly useful form
of the identity equivalent to $E_{7(7)}$ current conservation.

 The new developments  suggest that we may need to reconsider the predictions of  $E_{7(7)}$. The ``counterterm wisdom''  was that  the  counterterms should preserve the symmetries of the classical action. Also, the  counterterms which vanish on shell may be removable by a gauge choice.  For example, the counterterms in pure gravity depend on the Riemann-Chistoffel tensor $R_{\mu\nu\lambda \delta}$ and its covariant derivatives. The 2-loop counterterm of the form $R_{\mu\nu\lambda \delta} R^{\lambda \delta \kappa\eta} R_{\kappa\eta}{}^{ \mu\nu}$ can be added to the classical Lagrangian without affecting  the symmetries of the theory: the Lagrangian with the counterterms
 \be
S_{\rm deformed}= S_{0}+ S_{\rm CT}= \int \sqrt {-g }\left ( {1\over 2\kappa^2} R+ a \kappa^2 R_{\mu\nu\lambda \delta} R^{\lambda \delta \kappa\eta} R_{\kappa\eta}{}^{ \mu\nu} +...\right )
 \ee
 remains invariant under the standard, undeformed general covariance transformations
\be
(\Delta g_{\mu\nu})_0  = (\Delta g_{\mu\nu} )_{\rm deformed} = D_\mu \xi_\nu (x)+D_\nu \xi_\mu (x) \ ,
\ee
 and is ready to absorb the UV divergences. Note that the presence of the counterterms in the deformed action does not require the deformation of the symmetry!  The situation with other gauge symmetries is very similar. When the classical action is deformed by counterterms,
 \be
 S_{\rm deformed}= S_0 + S_{\rm CT} \ ,
 \ee
 it remains invariant under the undeformed gauge symmetries
$
 \Delta_0^{\rm gauge}  S_{\rm deformed}=  \Delta_0^{\rm gauge} (S_{0}  +  S_{\rm CT}) =0.$
{\it  The situation with continuous
global  duality symmetries  is more delicate, namely, even the classical Lagrangian is not invariant under  duality symmetry, only equations of motion are}.\footnote{ The prototype of electric-magnetic duality is the free Maxwell theory: in vacuum equations of motion and Bianchi identities are  invariant under rotations.  The Maxwell Lagrangian, $E^2-B^2$, is not invariant under $E \Leftrightarrow B$ rotation, only the Hamiltonian, $E^2+B^2$, is invariant, and the Noether-Gaillard-Zumino current is conserved.}
 The current consists of the standard Noether contribution for all fields but vectors, $j_\mu={\partial L\over \partial \varphi_{, \mu}} \Delta \varphi$,  and of the  Gaillard-Zumino current for vector fields, $\hat J_\mu$,  so that the total current  is conserved,
\be
\partial _\mu J^\mu_{NGZ}= \partial _\mu j_\mu + \partial _\mu \hat J_\mu=0 \ .
\ee
The fact that the Lagrangian is not invariant under duality leads to a dramatic difference with regard to the properties of the admissible counterterms, assuming that the corresponding duality symmetry has no anomalies. The difference comes from the properties of the transformations of the deformed Lagrangian including the classical part and the counterterms.    In addition to the field strength  $F=dA$ present in the action, the symmetry also involves the dual field strength defined by the variation of the action over $F$, namely $\tilde G = 2{\delta S\over \delta F}$. This is a basic unitarity issue. For example, in ${\cal{N}}=8$ supergravity there are 28 real fields, but  the $E_{7(7)}$ symmetry, mixing 28 Bianchi identities and 28 field equations of motion, requires a 56-dimensional doublet of fields $(F, G)$, where 28 $G$'s  depend on 28 $F$'s  and on scalars. The symmetry is
\be
\Delta \tilde F= A \tilde F + B \tilde G=  A \tilde F + 2 B {\delta (S_0 + S_{\rm CT})\over \delta F}= \Delta_0 \tilde F+B {\delta S_{\rm CT}\over \delta F}\, ,
\ee
where $A,B$ are infinitesimal parameters of transformation. The ${\cal{N}}=8$ supergravity
 counterterms constructed in  \cite{Howe:1980th,Kallosh:1980fi} are  $E_{7(7)}$ invariant under the undeformed symmetry  $\Delta_0 $ associated with the classical action. The counterterms depend on $F$,  therefore $ {\delta S_{\rm CT}\over \delta F}\neq 0$, which deforms $E_{7(7)}$  symmetry. In ${\cal{N}}=8$ supergravity $B= \mbox{Im}\Lambda+\mbox{Im}\Sigma$ are the off diagonal components of duality symmetries, which mix  Bianchi identities with equations of motion:
  \begin{eqnarray}
\Delta \,  \partial_{\mu} \tilde F^{\mu\nu} =B \,
                  \partial_{\mu} \tilde G^{\mu\nu} \ .
                      \end{eqnarray}
 Therefore
the Lagrangian  including the counterterms does not automatically lead to  the $E_{7(7)}$ symmetric deformed equations of motion. We have to find out whether the $E_{7(7)}$ symmetry  deformed by the counterterms remains valid. This means that we have to find out whether equations of motion/Bianchi identities transform into each other by the deformed $E_{7(7)}$   transformations, and in this way to  find the implications of the continuous global non-compact $E_{7(7)}$  symmetry for the quantum theory.

  In Sec. 2 we introduce the concept of a duality doublet $(F, G)$ which has a double amount of field strengths as compared to the one present in the action: $G$ is a functional of $F$ and other fields, or vice versa. We use the example of classical ${\cal{N}}> 2$ supergravity to explain it.
  In Sec. 3 we  discuss duality symmetry and  Noether-Gaillard-Zumino identity \cite{Gaillard:1981rj,Aschieri:2008ns},  for any symmetry which may be embedded into $Sp(2n,{\mathbb{R}})$. We derive a particular form of the duality identity for the case of ${\cal{N}}=8$ supergravity, where
the counterterms are invariant under undeformed duality.
 In Sec. 4 we analyze the 3-loop and higher loop counterterms of ${\cal{N}}=8$ supergravity, compute the deformation of the dual field strength caused by the counterterms and show that the deformation breaks $E_{7(7)}$ current conservation identity. In Sec. 5 we argue that the unitarity constraint which allows to express half of the 56-dimensional $E_{7(7)}$ doublet via the independent 28 fields is unique. This leads to an independent argument that the hidden $E_{7(7)}$ symmetry combined with all manifest gauge symmetries of ${\cal{N}}=8$ supergravity forbids all  counterterms in ${\cal{N}}=8$ supergravity. In Sec. 6 we discuss our findings and provide a short technical summary of the results. Appendix A contains a derivation of the NGZ identity and simple expressions for the scalar and vector part of the $E_{7(7)}$ current of the $SU(8)$ locally invariant ${\cal{N}}=8$ supergravity.  In Appendix B we discuss the Hamiltonian approach to dualities.

 \section{Duality doublet  $(F, G)$}
 \setcounter{equation}{0}
\subsection{Classical ${\cal{N}}> 2 $ supergravity}

Here we start with the classical $d=4$  ${\cal{N}}> 2$ pure supergravity and explain its  duality symmetry.\footnote {We use notation of \cite{Aschieri:2008ns}, including $\tilde G^{\mu\nu}= \sqrt g \, G^{*\mu\nu}$ where the Hodge dual of a two form is $\Omega_{\mu\nu}^*= {1\over 2} \sqrt g \epsilon_{\mu \nu \lambda \sigma}   \Omega^{\lambda \sigma}$. In the flat case $\tilde G^{\mu\nu}= {1\over 2} \epsilon^{\mu\nu\lambda \sigma} G_{\lambda \sigma}$ and $G_{\mu\nu}= -{1\over 2} \epsilon_{\mu\nu\lambda \sigma} \tilde G^{\lambda \sigma}$ ($\epsilon ^{0123}= -\epsilon_{0123}=1$).
We will also suppress spacetime indices so that for example $F\tilde G= F_{\mu\nu} \tilde G^{\mu\nu}$. It is useful to remember that $F\tilde G= \tilde F G$ and $\tilde{\tilde F}  = -F$, $\tilde F\tilde G= -FG$ where $FG= F^{\mu\nu} G_{\mu\nu}$.}
In absence of fermions four-dimensional supergravities depend on metric, vectors and scalars. In particular, the action depends on  Abelian vectors ${\cal A}_\mu^{\Lambda}$ via the field strength $F_{\mu\nu}^{\Lambda}= \partial_\mu {\cal A}_\nu^{\Lambda}-\partial_\nu {\cal A}_\mu^{\Lambda}$, on scalars and on metric
\be
S_{\rm cl}(F, \phi, g)= {1\over 4 \kappa^2} \int d^4x \, e \Big( -{1\over 2} R+ {\rm Im} {\cal N}_{\Lambda \Sigma}
F_{\mu\nu}^\Lambda F^{\mu\nu \Sigma}  + {1\over 2 \, e} {\rm Re} {\cal N}_{\Lambda \Sigma} \epsilon^{\mu\nu\rho\sigma} F_{\mu\nu}^\Lambda F_{\rho\sigma}^\Sigma + {1\over 2} g_{ij}(\phi) \partial_\mu \phi^i \partial_\mu \phi^j \Big )  .
\label{0}\ee
Here the kinetic terms for vectors ${\cal N}_{\Lambda \Sigma}(\phi)$ depends on scalars. The manifold of scalars  is a coset space ${G/H}$. In case of ${\cal{N}}=8$ supergravity  ${G/H}= E_{7(7)}/SU(8)$.
There is a Bianchi identity for  $F_{\mu\nu}^{\Lambda}= \partial_\mu {\cal A}_\nu^{\Lambda}-\partial_\nu {\cal A}_\mu^{\Lambda}$,
\be
\partial_\mu \tilde F^{\mu\nu \, \Lambda }=0\, .
\ee
The  dual field strength $G^{\mu\nu}_\Lambda $ is defined as a derivative of the action over $F_{\mu\nu} $,  namely
$
\tilde G^{\mu\nu}_{\Lambda }=2 {\delta S (F, \phi, g)\over \delta  F_{\mu\nu}^\Lambda}
$.
Equations of motion provide the Bianchi identity for the dual field strength
\be
 \partial_\mu {\delta S \over \delta F_{\mu\nu}}=0 \qquad \Rightarrow \qquad \partial_\mu \tilde G^{\mu\nu } =0   \qquad \Rightarrow \qquad G_{\mu\nu } = \partial_\mu {\cal B}_{\nu }-\partial_\nu {\cal B}_{\mu } \ .
\label{dual1} \ee
The dual potential ${\cal B}_\mu$ exists only when the equation of motion of the deformed theory are satisfied.
One can evaluate $G$ using the action in (\ref{0}) and one finds in absence of fermions\footnote{When fermions are present in classical supergravity, $\tilde G$ has also terms independent on $F$.}
\be
\tilde G^{\mu\nu}_{\Lambda }=2 {\delta S\over \delta  F_{\mu\nu}^\Lambda}= {1\over \kappa^2}\Big ( e\, {\rm Im} {\cal N}_{\Lambda \Sigma}
F^{\mu\nu \Sigma}+ {1\over 2 } {\rm Re} {\cal N}_{\Lambda \Sigma} \epsilon^{\mu\nu\rho\sigma}  F_{\rho\sigma}^\Sigma \Big )= \tilde G^{\mu\nu}_{\Lambda } (F, \phi, g) \ .
\label{G0}\ee
For ${\cal{N}}> 2$ supergravity duality symmetry  can be embedded into $Sp(2n,{\mathbb{R}})$.
It requires that the vector doublet $(F, G=2 {\delta S\over \delta F})$ transforms in the fundamental of $Sp(2n,{\mathbb{R}})$
\begin{eqnarray}\label{symplecticBig}
\left(
                      \begin{array}{cc}
                  F' \\
                    G' \\
                      \end{array}
                    \right)\ ={\cal S}  \left(
                      \begin{array}{cc}
                  F \\
                    G \\
                      \end{array}
                    \right) \ ,  \qquad  \qquad {\cal S} \equiv \left(
                      \begin{array}{cc}
                      \hat A& \hat B \\
                       \hat C & \hat D \\
                      \end{array}
                    \right) \ .
                    \end{eqnarray}
Here   the matrix ${\mathcal{S}}$ is symplectic\footnote{We will later use the infinitesimal form of duality symmetry. This will correspond to
$
\hat A\approx 1+A$,  $\hat B\approx B$,   $\hat C\approx C$,   $\hat D\approx 1+D$.}, it has real elements that satisfy the following conditions: $
\hat A^{\mathsf{T}} \hat C-\hat C^{\mathsf{T}} \hat A= \hat B^{\mathsf{T}}\hat D- \hat D^{\mathsf{T}}\hat B=0\ $ and $ \hat A^{\mathsf{T}}\hat D- \hat C^{\mathsf{T}}\hat B=1$.
The gauge kinetic term ${\cal N}$ transforms via fractional transformations
\be
{\cal N}'=(\hat C+ \hat D {\cal N})( \hat A + \hat B {\cal N})^{-1} \ .
\label{sctransf}\ee
Note that the duality (\ref{symplecticBig}) requires that
\be
G'= \hat CF + \hat D G  \qquad \Rightarrow  \qquad \tilde G'= \hat C\tilde F + \hat D \tilde G \ .
\label{G0'}\ee
On the other hand one can evaluate $G'$ using its  expression in (\ref{G0}) where it depends on $F$ and scalars and using that from (\ref{symplecticBig})
$
F'= \hat A F + \hat B G \ ,
$
as well as the transformations of scalars in (\ref{sctransf}). This means that the transformation of the dual field strength $G(F, \phi, g)$ computed via its relation to $F, \phi, g$ using the chain rule,  gives the same answer as the one required by the symmetry in the form (\ref{symplecticBig}).
The fact that classical supergravities have duality symmetries with the conserved current means that  $\tilde G'$ in (\ref{G0'})  coincides with the expression following from the constraint between $G$ and $F, \phi$ in (\ref{G0}).
This is the essence of Noether-Gaillard-Zumino $Sp(2n,{\mathbb{R}})$ construction for duality invariant theories.

When counterterms are added to the action  $F_{\mu\nu}^{\Lambda}$, of course, remains undeformed, $F_{\mu\nu}^{\Lambda}= \partial_\mu {\cal A}_\nu^{\Lambda}-\partial_\nu {\cal A}_\mu^{\Lambda}$.  However the dual field strength $G_{\mu\nu}$  is deformed when the action is deformed
\be
 \tilde G^{\mu\nu\, \rm deformed} (F, \phi, g)
 = 2 {\delta S_0 \over \delta F_{\mu\nu}} (F, \phi, g)+2 {\delta S_{\rm CT} \over \delta F_{\mu\nu}} (F, \phi, g) \ .
 \label{dual} \ee
The problem we will address with regard to a deformed action $S_0+ S_{\rm CT}$ is: given the explicit dependence of $ \tilde G^{\mu\nu\, \rm deformed}$ on $(F, \phi, g)$,  which we will compute from the $S_{\rm CT}$, will it provide the  deformed duality transformation corresponding to the doublet transformations  of $(F, G^{\rm deformed})$?

   \section{ Noether-Gaillard-Zumino  (NGZ) $Sp(2n,{\mathbb{R}})$ duality identitiy}
 \setcounter{equation}{0}
 Here we review the NGZ construction \cite{Gaillard:1981rj,Aschieri:2008ns} for theories with $Sp(2n,{\mathbb{R}})$ duality,
which has some real vector fields ${\cal A}_\mu^\Lambda$ and other fields $\varphi^\alpha$, which include scalars, spinors, metric. There is an infinitesimal $Sp(2n,{\mathbb{R}})$ transformation, which acts on $Sp(2n,{\mathbb{R}})$ doublet of vectors field strength $   (F,G)
                   $ as follows:
\begin{eqnarray}\label{symplectic}
\Delta \left(
                      \begin{array}{cc}
                  F \\
                  G \\
                      \end{array}
                    \right)\ =\left(
                      \begin{array}{cc}
                      A&  B \\
                        C & D \\
                      \end{array}
                    \right)  \left(
                      \begin{array}{cc}
                  F \\
                    G \\
                      \end{array}
                    \right) \ , \qquad \Delta \left(
                      \begin{array}{cc}
                  \tilde F \\
                  \tilde G \\
                      \end{array}
                    \right)\ =\left(
                      \begin{array}{cc}
                      A&  B \\
                        C & D \\
                      \end{array}
                    \right)  \left(
                      \begin{array}{cc}
                  \tilde F \\
                    \tilde G \\
                      \end{array}
                    \right) \ .
\end{eqnarray}
\begin{eqnarray}\label{infinitesimal}
 A^{\mathsf{T}} =-D \ ,  \qquad B^{\mathsf{T}}= B \ ,  \qquad C^{\mathsf{T}}=C \ .
\end{eqnarray}
 Here $A,   B,  C, D$ are the infinitesimal global (space-time independent) parameters of the transformations, arbitrary real $n\times n$ matrices satisfying (\ref{infinitesimal}). It is an infinitesimal version of (\ref{symplecticBig}).
 The upper component of the doublet $F_{\mu\nu}^{\Lambda}= \partial_\mu {\cal A}_\nu^{\Lambda}-\partial_\nu {\cal A}_\mu^{\Lambda}$ is the field strength of the vector field, the down component is a dual field strength,
\be\tilde G^{\mu\nu}_{\Lambda}[F, \varphi]  \equiv 2  {\delta S[F, \varphi]\over \delta F_{\mu\nu}^\Lambda} \ .
\ee
   Duality symmetry on $\varphi^\alpha$ fields is of the form
\be
\Delta \varphi^\alpha= \Xi^\alpha(\varphi) \ .
\label{chi}\ee
There is a consistency requirement here that the dual field strength  transforms according to  (\ref{symplectic}) using the chain rule, when expressed as a functional of $F$ and $\varphi$. This consistency condition\footnote{ It is equivalent to the Noether current conservation. The current has some unusual properties, see Appendix B.  }  is given in the form of Noether-Gaillard-Zumino  $Sp(2n,{\mathbb{R}})$ duality identify which we will present and use below.
If the identity is satisfied,
duality symmetry of equations of motion follows from (\ref{symplectic}) since it mixes Bianchi identities with the equations of motion:
\begin{eqnarray}\label{symplecticdual}
\Delta \left(
                      \begin{array}{cc}
                  \partial_{\mu} \tilde F^{\mu\nu} \\
                  \partial_{\mu} \tilde G^{\mu\nu} \\
                      \end{array}
                    \right)\ =\left(
                      \begin{array}{cc}
                      A&  B \\
                        C & D \\
                      \end{array}
                    \right)  \left(
                      \begin{array}{cc}
                  \partial_{\mu} \tilde F^{\mu\nu} \\
                  \partial_{\mu} \tilde G^{\mu\nu} \\
                      \end{array}
                    \right) \ .
\end{eqnarray}
 A consistency of the $Sp(2n,{\mathbb{R}})$  duality symmetry requires that {\it the Lagrangian must transform under duality in a certain way, defined by  NGZ identitiy}
  \cite{Gaillard:1981rj,Aschieri:2008ns}
 \be\boxed{
 {\delta \over \delta F^\Lambda }\Big (  S[F', \varphi']- S[F, \varphi]- {1\over 4} \int (\tilde FCF+ \tilde GBG) \Big)=0.}\label{GZ}
 \ee
 This identity is suitable for the situation  that the action depends  on $F_{\mu\nu}$ and on its derivatives. Instead of partial Lagrangian derivatives  ${\partial L\over \partial  F^\Lambda }$ one has to use the variational derivatives of the action, ${\delta S\over \delta F^\Lambda }$, as suggested in (\ref{GZ}).  We present a derivation of the  identity  (\ref{GZ}) in Appendix A.

\subsection{Using the symmetry of counterterms under undeformed duality}
All deformations will be denoted by the hat symbol. \be
S=S_0+ \hat S \ , \qquad    G=G_0 + \hat G \ , \qquad \tilde G=\tilde G_0 + \hat {\tilde G}\ , \qquad \Delta = \Delta_0  + \hat \Delta \ ,
\ee
$F$ is not deformed, $\Delta \varphi^\alpha$ is not deformed.
\be
\Delta S  =  \Delta_0 S_0 + \Delta_0 \hat S+
\hat \Delta S_0+  \hat \Delta \hat S \ .
\ee
If
the counterterms are invariant under undeformed duality, like in  ${\cal{N}}=8$ supergravity, it means that
\be
 \Delta_0  S_{\rm CT} = \Delta_0  \hat S = 0 \ .
\ee
The  NGZ identity (\ref{GZ})
takes the form
\be
 {\delta \over \delta F^\Lambda } \Big (  \Delta_0 S_0 +
\hat \Delta S_0+  \hat \Delta \hat S
- {1\over 4} \int (\tilde FCF+ \tilde GBG)\Big )=0 \ .
\ee
The undeformed symmetry of the undeformed action cancels  in this expression since
\be
 {\delta \over \delta F^\Lambda } \Big (  \Delta_0 S_0
- {1\over 4} \int (\tilde FCF+ \tilde G_0BG_0)\Big )=0 \ .
\ee
and the remaining identity is (taking into account that $GB\tilde G= (G_0+\hat G)B (\tilde G_0+ \hat{\tilde  G})$)
\be
 {\delta \over \delta F^\Lambda } \Big ( \hat \Delta S
- {1\over 4} \int ( 2 \tilde G_0B\hat G + \hat{\tilde  G} B\hat G )\Big )=0 \ .
\label{rem}\ee
Note that the deformation of duality  enters only via the deformation of the symmetry on vectors
\be
\hat \Delta S = {\delta S \over \delta F} B \hat G= {1\over 2} \tilde G B \hat G ={1\over 2} \tilde G_0 B \hat G+{1\over 2} \hat {\tilde G} B \hat G \ .
\ee
 We plug this back to NGZ identity (\ref{rem}) and find that
\be
 {\delta \over \delta F^\Lambda }\int  \Big ( {1\over 2} \tilde G_0 B \hat G+ {1\over 2} \hat {\tilde G} B \hat G
-  {1\over 4}  ( 2 \tilde G_0B\hat G + \hat{\tilde  G} B\hat G )\Big )=0 \ ,
\ee
which requires that
\be\boxed{
 {\delta \over \delta F^\Lambda }\int  \Big (\hat {\tilde G} B \hat G\Big ) =0.  }
\label{critical}\ee
This puts a strong restriction on the deformation of the action $\hat S$ and the possible form of the deformation of the dual field strength $\hat G$.

 \section{${\cal{N}}=8$ Supergravity Counterterms and $E_{7(7)}$ Identity}
 \setcounter{equation}{0}
The on shell superspace \cite{Brink:1979nt} provides a geometric construction  of counterterms \cite{Howe:1980th,Kallosh:1980fi} where the torsion and curvature superspace tensors $T^M_{KL}$ and $R_{MNKL}$ are manifestly Lorentz and $SU(8)$  covariant.
\be
L_{\rm CT}= L_{\rm CT} \Big (T^P_{KL}(x, \theta) , R_{PQKL} (x, \theta)\Big ) \ .
\label{CT}\ee
These tangent space  tensors transforms as tensors under the local Lorentz and local $SU(8)$ transformations and they are neutral under the classical, undeformed $E_{7(7)}$, it is hidden. On shell means that every superfield satisfies a non-linear classical equation of motion.
The counterterms depend on $E_{7(7)}$ non-covariant field strength $F_{\mu\nu}^{IJ}$ and on scalars ${\cal V}$,  but only in a combination which does not transform on undeformed $E_{7(7)}$.

\noindent 1. {\it Vectors}:  The $E_{7(7)}$ vector doublet in classical on shell ${\cal{N}}=8$ supergravity is defined as follows:
\be
d({\cal X}_{IJ}^0, \bar {\cal X}^{0\, IJ })=0 \ ,
\ee
which means that on shell there 28 complex or  56 real potentials: 28 ${\cal A}^{IJ}$ in the classical action and another 28 ${\cal B}_{IJ}^0$ are dual
\be
{\cal X}_{IJ}^0= d ({\cal B}_{IJ}^0+ i{\cal A}^{IJ}) \, , \qquad \bar {\cal X}^{0 IJ}= d ({\cal B}_{IJ}^0- i{\cal A}^{IJ}) \ .\ee
Here ${\cal A }^{IJ} = dx^\mu {\cal A}_\mu ^{IJ}$  and ${\cal B} _{IJ} = dx^\mu {\cal B}_{\mu IJ}$ are real and
\be
F_{\mu\nu}^{IJ}= \partial_\mu  {\cal A}_\nu ^{IJ }-  \partial_\nu  {\cal A}_\mu ^{IJ}\,  , \qquad G_{\mu\nu IJ}^0= \partial_\mu  {\cal B}_{\nu IJ}^0-  \partial_\nu  {\cal B}_{\mu IJ}^0= 2{\partial S_0[F, \phi]\over \partial  F_{\mu\nu}^{IJ}} \ .
\ee
The spin-one field strengths which transforms as a doublet under  $SU(8)$ depend on the $U(1)$ field strength  $E_{7(7)}$ doublet $({\cal X}_{IJ}^0, \bar {\cal X}^{0\, IJ })$  and on scalars ${\cal V}$ as follows
\be
({\cal F}_{ij} ^0, \; \overline {\cal F}^{0\, ij})= ({\cal X}_{IJ}^0 , \; \overline {\cal X}^{0\, IJ}) {\cal V} \ .
\label{BrinkHowe}\ee
Here $ij$ are the $SU(8)$ and $IJ$ are the $E_{7(7)}$ indices and
\be
{\cal V}=\left  (\begin{array}{cc}
                      U^{IJ}{}_{ij}&  \bar V^{IJij} \\
                      V_{IJij} & \bar U_{IJ}{}^{ij} \\
                      \end{array}\right) \ .
                    \label{V}
                    \ee
 is a vielbein-like object describing the 133 scalars of ${\cal{N}}
 =8$ supergravity with an unbroken local $SU(8)$. The corresponding coset space geometry is ${E_{7(7)}\over SU(8)}$. When local  $SU(8)$ symmetry is gauge-fixed, for example in the unitary gauge, ${\cal V}={\cal V}^\dagger$, only  70 physical scalars remain: $\phi_{ijkl}= {1\over 4!} \epsilon_{ijlkmnpq} \bar \phi^{mnpq}$.   There is no difference between $IJ$ and $ij$ indices anymore and only global  $SU(8)$ remains as the symmetry of the action, whereas  the global $E_{7(7)}$ is the symmetry mixing equations of motion with Bianchi identities. The non-linear $E_{7(7)}$ symmetry acts on all (but metric) fields of the theory in ${\cal V}={\cal V}^\dagger$ gauge, the compact form of the  transformations   is given in \cite{Kallosh:2008ic}.

The counterterms in \cite{Howe:1980th,Kallosh:1980fi} are known in the form of the theory with local not gauge-fixed $SU(8)$ symmetry. This simplifies the analysis of duality, where ${\cal V}$ is a 133-component group element of $E_{7(7)}$
 which transforms under local $SU(8)$  and by a global $E_{7(7)}$ symmetry: the global $E_{7(7)}$ acts on the capital indices  and
the local $SU(8)$ acts on the lower case indices
\begin{eqnarray}\label{V2}
{\cal V}' =  E^{-1} {\cal V} U(x) \, ,
\qquad
({\cal X}_{IJ}^0 , \; \overline {\cal X}^{0\, IJ})' = ({\cal X}_{IJ}^0 , \; \overline {\cal X}^{0\, IJ})  E \ .
\label{classical}\end{eqnarray}
\begin{eqnarray}\label{V1}
E =e^{ G_{_{E_{7(7)}}}}\, ,  \qquad G_{_{E_{7(7)}}} ={\left(
         \begin{array}{cc}
           2 \delta^{[I}_{[K} \Lambda^{J]}{}_{L]} &  \bar \Sigma^{IJKL} \\
           {\Sigma}_{IJKL} & 2 \delta^{[K}_{[I} \Lambda^{L]}_{}{J]} \\
         \end{array}
       \right)}\, ,
\label{classical}\end{eqnarray}
\begin{eqnarray}\label{gauge}
U(x) = \exp G_{SU(8)} \ ,   \qquad
G_{SU(8)}(x) =\left(
                                        \begin{array}{cc}
                                         \delta^{[i}{}_{[k} \Lambda^{j]}{}_{l]}(x) & 0 \\
                                          0& \delta_{[m}{}^{[p} \Lambda_{n]}{}^{q]}(x)\\
                                        \end{array}
                                      \right)\ .
\end{eqnarray}
The undeformed infinitesimal $E_{7(7)}$ symmetry acts on classical vector doublets in the real basis as follows \cite{Kallosh:2008ic}
\begin{eqnarray}\label{deltaFG}
   \Delta\left(
  \begin{array}{c}
  F \\
 G^0 \\
  \end{array}
\right)=\left(
          \begin{array}{cc}
             \mbox{Re}\Lambda-\mbox{Re}\Sigma & \mbox{Im}\Lambda+\mbox{Im}\Sigma \\
            -\mbox{Im}\Lambda+\mbox{Im}\Sigma & \mbox{Re}\Lambda+\mbox{Re}\Sigma \\
          \end{array}
        \right)\left(
                  \begin{array}{c}
                    F \\
                    G^0 \\
                  \end{array}
                \right)\ .
\end{eqnarray}
The $SU(8)$ doublet is now constructed from the $E_{7(7)}$ doublet and the vielbein as shown in (\ref{BrinkHowe}). The field strength ${\cal F}_{ij}^0 $  ($\overline {\cal F}^{0\, ij} $)  transforms as a $\underline{\mathbf{28 }}$ (${\mathbf{28 }}$) under $SU(8)$ and it is invariant under $E_{7(7)}$:
\be
({\cal F}_{ij}^0 , \; \overline {\cal F}^{0\, ij})'=   ({\cal X}_{IJ}^0 , \; \overline {\cal X}^{0\, IJ})' { \cal V}'= ({\cal X}_{IJ}^0 , \; \overline {\cal X}^{0\, IJ}) E \, E ^{-1}  {\cal V}  U(x)=({\cal F}_{ij}^0 , \; \overline {\cal F}^{0\, ij})U(x) \ .
\label{BrinkHowe1}\ee
The field strength ${\cal F}_{ij}$  transforms as a $\underline{\mathbf{28 }}$ under  $SU(8)$ and  is space-time complex self-dual
\be
{\cal F} _{ij \mu\nu}^0= {1\over 2} (\sigma_{\mu\nu}) ^{\alpha \beta} M_{\alpha \beta ij}\, ,  \qquad  {\cal F} _{ij \mu\nu}^{* 0}= i {\cal F} _{ij \mu\nu}^0 \ .
\label{constr0}\ee
Counterterms depend on vector fields  only via $M_{\alpha \beta ij}$ and its conjugate.
Equation (\ref{BrinkHowe}) together with the constraint (\ref{constr0}) allow to find the relation between $F$ and $G^0$. It is the same relation which follows from the classical action and definition $G^0=2{\delta S_0\over \delta F}$. One finds that in absence of fermions
\be
G_{ \dot \alpha \dot \beta}^0= - i F_{ \dot \alpha \dot \beta}  (U-V)( U+V)^{-1} \ . \label{constr}\ee
This relation is the same as the one in (\ref{G0}) for ${\cal{N}}=8$ supergravity.

\noindent 2. {\it Scalars}: the scalars enter in counterterms only via a combination which  transform under $SU(8)$ but is neutral under $E_{7(7)}$. Namely, the $SU(8)$ covariant derivative of the vielbein is contracted with the inverse vielbein
\be
{\cal V}^{-1} D {\cal V}= {\cal V}^{-1} d {\cal V} + \left  (\begin{array}{cc}
                     Q&  0\\
                     0 & \bar Q \\
                      \end{array}\right)=
\left  (\begin{array}{cc}
                     0&  \bar P \\
                      P & 0 \\
                      \end{array}\right) \ .
\ee
The counterterms depend on $SU(8)$ tensors
$
P_{ijkl}$  and $  \bar P^{ijkl}$,  and $ \bar P^{ijkl}= {1\over 4!} \epsilon^{ijklmnpq} P_{mnpq}$. When ${\cal V}$ transforms under $E_{7(7)}$  and  $SU(8)$ as
$
{\cal V}' =  E^{-1} {\cal V} U(x)
$ the scalar combination in
${\cal V}^{-1} D {\cal V}$ is  $SU(8)$ covariant and $E_{7(7)}$ invariant
\be
{\cal V'}^{-1} D' {\cal V'}= U(x)^{-1} {\cal V}^{-1} D {\cal V} U(x) \ .
\ee
The linearized version of $P_{\alpha \dot \beta \, ijkl}$ is a derivative of the scalar field, $P_{\alpha \dot \beta \, ijkl}=\partial_{\alpha \dot \beta} \phi_{ijkl}+...$ and $\bar P_{\alpha \dot \beta}^{  ijkl}=\partial_{\alpha \dot \beta} \bar \phi^{ijkl}+...$.

The torsion and curvature superspace tensors depend on $M_{\alpha \beta ij}$ and its $SU(8)$ conjugate $\bar M^{\alpha \beta ij}$ and on $P_{\alpha \dot \beta \, ijkl}$, $\bar P^{ijkl}_{\alpha \dot \beta}$ and their $SU(8)$ supercovariant derivatives. The counterterms are invariant under undeformed $E_{7(7)}$ symmetry since they depend only on $SU(8)$ tensors constructed from $F$ and $G^0=2{\partial S_0\over \partial  F}$.  When the action is deformed by the counterterms, the dual field strength is modified
\be
G^0 \rightarrow G^0 + \hat G  \ .
\ee
Consistency of the deformed duality symmetry requires that  ${\delta \over \delta F^\Lambda }\int  \Big (\hat {\tilde G} B \hat G\Big ) $  vanish. Here 70 parameters $B$ are given by $( \mbox{Im}\Lambda+\mbox{Im}\Sigma)$.

\subsection{  $\hat G$ from  counterterms and $E_{7(7)}$ identity}
The counterterms in ${\cal{N}}=8$ supergravity  \cite{Howe:1980th,Kallosh:1980fi} depend on   $ F_{\mu\nu}$,  $D_\lambda F_{\mu\nu}$,  $D_\delta D_\lambda F_{\mu\nu}$ etc. For example, the linearized 3-loop counterterm has terms linear, quadratic and quartic in $F$ and its derivatives.
The 3-loop linearized candidate counterterm  was presented in \cite{Kallosh:1980fi} as an integral over 16 Grassmann variables. The corresponding  superaction is
\be
S^{3loop}\sim  \,   \kappa^4 \int d^4x d^{16}\theta_B \, W_{1234}^4(x, \theta)\sim  \, \kappa^4 \int d^4x \Big ( R^4+ (\partial F)^2 R^2+...\Big ) \ .
\label{3loop}\ee
To get all component expressions one has to perform  16 $\theta$ integration.
Some of these terms were identified in \cite{Kallosh:2008ru} using helicity formalism of the amplitudes.
More recently the candidate 3-loop counterterm
was  presented in \cite{Freedman:2011uc} in components, with {\it 51 explicit quartic monomials} depending on all component fields of the theory.  Of these {\it 29 depend $F$ and its derivatives}.
The 7-loop candidate counterterm  has analogous dependence on $F$, just more derivatives. At the linear level in momentum space it has an extra factor $(s^2+ t^2+u^2)^2$ in terms of Mandelstam variables
\be
S^{7loop}\sim  \,  \kappa^{12}  \int  d^{16} \theta_B \prod _{i=1}^4 d^4 p_i   \, W_{1234}(p_i, \theta) (s^2+ t^2+u^2)^2 \ .
\label{7loop}\ee

If the computation in \cite{Bern:2007hh} would not show that the theory is UV finite in 3-loops,   we would have to add  $S^{3loop}$ to the classical action so that the UV divergence can be absorbed.
To check the status of NGZ identity for the action $S_0+ S^{3loop}$ we pick up one (from 51) term in eq. (6.8) in \cite{Freedman:2011uc}.  We focus on $ (\partial F)^2 R^2$ term which in 2-component notation is given by
\be
S^{3loop}_{ _{(\partial F)^2 R^2}}
\sim x R_{\dot \alpha \dot \beta \dot \gamma\dot \delta} F^{\dot \alpha \dot \beta ij } \partial^{\dot \gamma \gamma} \partial^{\dot \delta \delta} F^{\alpha \beta}_{ij} R_{ \alpha  \beta  \gamma \delta} \ .
\ee
Here $F^{\dot \alpha \dot \beta ij }$ and its conjugate $F^{\alpha \beta}_{ij}$ may be viewed as linearized $SU(8)$ tensors  which are neutral under classical $E_{7(7)}$ duality symmetry. They are related to $E_{7(7)}$ doublet at the non-linear level, as shown in (\ref{BrinkHowe}). In the linear approximation $F^{\alpha \beta}_{ij}\approx F^{\alpha \beta}_{ IJ}$ and $F^{\dot \alpha \dot \beta ij }\approx F^{\dot \alpha \dot \beta IJ }$. This means that the deformation of the dual field strength caused by the $(\partial F)^2 R^2$  part of the counterterm is
\be
\hat G_{\dot \alpha \dot \beta IJ } \sim {\delta S^{3loop}_{ (\partial F)^2 R^2}\over \delta F_{\dot \alpha \dot \beta}^{IJ} }=  x R_{\dot \alpha \dot \beta \dot \gamma\dot \delta}  \partial^{\dot \gamma \gamma} \partial^{\dot \delta \delta} F^{\alpha \beta}_{IJ} R_{ \alpha  \beta  \gamma \delta} \ .
\label{hatG} \ee
Now we have to test the $E_{7(7)}$ identity in the form (\ref{critical}) which in 2-component notation requires that
\be
{\delta \over \delta F_{\alpha \beta}^{MN}}\int \Big ( \hat G_{\dot \alpha \dot \beta IJ} B^{IJKL} \hat G^{\dot \alpha \dot \beta}_{ KL} -h.c.\Big )=0 \ .
\label{critical0}\ee
Substituting (\ref{hatG}) into (\ref{critical1}) we find that
\be
x^2 R_{\dot \alpha \dot \beta \dot \gamma\dot \delta}  \partial^{\dot \gamma \gamma} \partial^{\dot \delta \delta} \Big ( R_{ \alpha  \beta  \gamma \delta} B^{IJKL} \epsilon^{\dot \alpha \dot \alpha_1} \epsilon^{\dot \beta \dot \beta_1}R_{\dot \alpha_1 \dot \beta_1 \dot \gamma_1\dot \delta_1}  \partial^{\dot \gamma_1 \gamma_1} \partial^{\dot \delta_1 \delta_1} F^{\alpha_1 \beta_1}_{KL} R_{ \alpha_1  \beta_1  \gamma_1 \delta_1}\Big )=0 \ .
\label{proof}\ee
For generic curvatures and vectors (even with the account of  the linearized equations of motion $\partial^{\alpha \dot \alpha} R_{\dot \alpha \dot \beta \dot \gamma\dot \delta} =0$ and $\partial^{\alpha \dot \alpha} F_{\dot \alpha \dot \beta} =0$) the $E_{7(7)}$ identity (\ref{critical0}) is violated unless $x=0$.

There is no other term in 51 structures of the 3-loop counterterm in \cite{Freedman:2011uc} which can cancel $R^2\partial^2 R^2 \partial^2 F $ term in (\ref{critical0}),   as can be seen by a direct inspection.
The computations of the 3-loop UV divergence in \cite{Bern:2007hh} have shown that $x=0$. Equation (\ref{proof}) is an $E_{7(7)}$ symmetry prediction that $x$ has to vanish.

We could have focused on $(\partial F)^4$ term in the 3-loop counterterm, there are two such terms:
\be
S^{3loop}_{ _{(\partial F)^4}}
\sim x \Big(  F_{\dot \alpha \dot \beta}^{ ij } \partial_\mu \partial_\nu F^{\dot \alpha \dot \beta}_{ kl } \partial^\mu \partial^\nu F^{ \alpha \beta}_{ ij } F_{ \alpha  \beta kl }+ F_{\dot \alpha \dot \beta}^{ ij } \partial_\mu \partial_\nu F^{\dot \alpha \dot \beta}_{ kl } \partial^\mu  F^{ \alpha \beta}_{ ik } \partial^\nu F_{ \alpha  \beta jl }\Big ) \ .
\label{F4}\ee
The corresponding deformation of $\hat G_{\dot \alpha \dot \beta} $ due to these two terms would have 2 terms, from each term in (\ref{F4}). In the identity (\ref{critical0}) there will be $2\times 2+2\times 2=8$ terms since the second term in (\ref{critical0}) will also contribute. One would have to prove that all 8 terms do not cancel and this would require significantly more computations, as well as the use of various identities. It is therefore nice that in the sector of 2 gravitons and two vectors  we find only one contribution to the identity (\ref{critical0}) presented in (\ref{proof}). This expression is not vanishing unless $x=0$, i.e. the presence of the counterterm would break the $E_{7(7)}$ current conservation in the deformed theory.

We may now look at any exact $L$-loop counterterm in  \cite{Howe:1980th,Kallosh:1980fi} or in \cite{Beisert:2010jx} for the linearized form of these counterterms. At the level of a 4-point amplitude we will find terms like $\kappa^{2(L-1)} (\partial F)^2 \partial^{2(L-3)} R^2$. For example, the 7-loop counterterm will have $\kappa^{12} (\partial F)^2 R^2(s^2+t^2+u^2)^2$ terms as well as many other ones. The procedure of getting all required structures for the linearized partners of $D^{2k} R^4$ is described in \cite{Freedman:2011uc}.

The  computation of the deformation of the dual field strength to get an explicit expression for $\hat G$ becomes more involved since the number of terms with various distribution of extra derivatives grows and there will be more than one term to look at. However, there will be also growing number of structures in (\ref{critical1}) which all have to vanish.
We do not see any possibility to satisfy the identity (\ref{critical1}) in each sector, unless the coefficient in front of each counterterm vanishes.

 \section{Deformation of ${\cal{N}}=8$ supergravity}
  \setcounter{equation}{0}
 In this section we would like to find an alternative reason for the $E_{7(7)}$ current conservation forbidding  counterterms  constructed in \cite{Howe:1980th,Kallosh:1980fi}, which are compatible will all gauge symmetries and are invariant under undeformed $E_{7(7)}$.
We consider a possibility to deform the classical action  of ${\cal{N}}=8$ supergravity by the counterterms, which means that the deformed action  has to provide a 2-form doublet  of $E_{7(7)}$ such that the symmetry mixes vector equations with Bianchi identities:
\be
d({\cal X}_{IJ}, \bar {\cal X}^{IJ })=0 \ .
\ee
This means that on shell there are 28 complex (56 real) potentials,
$
{\cal X}_{IJ}= d ({\cal B}_{IJ}+ i{\cal A}^{IJ})$, $ \bar {\cal X}^{IJ}= d ({\cal B}_{IJ}- i{\cal A}^{IJ})$.
The corresponding double set of field strengths is given by
\be
F_{\mu\nu}^{IJ}= \partial_\mu  {\cal A}_\nu ^{IJ }-  \partial_\nu  {\cal A}_\mu ^{IJ}\,  , \qquad G_{\mu\nu IJ}= \partial_\mu  {\cal B}_{\nu IJ}-  \partial_\nu  {\cal B}_{\mu IJ} \ .
\ee
Only one of them shows up in the deformed action, the other one must be a functional of the first one, or vice versa, since
 there are only 28 real vectors in ${\cal{N}}=8$ supergravity.  So we need to find a relation between
${\cal X}_{IJ}$ and  $\bar {\cal X}^{IJ }$ (or $F^{IJ}$ and $G_{IJ}$) which picks up  28 dynamical degrees of freedom out of 56.
The corresponding constraint was discovered by Cremmer and Julia in the context of the  classical ${\cal{N}}=8$ supergravity \cite{Cremmer:1979up}. It requires as the first stage a  construction of the $SU(8)$ doublet.

To construct the  $SU(8)$ tensors one has to use the vielbein (\ref{V}).  Note that the relation between the $SU(8)$ doublet $({\cal F}_{ij} , \; \overline {\cal F}^{ij})$ and the $E_{7(7)}$ doublet $({\cal X}_{IJ} , \; \overline {\cal X}^{IJ})$ is unique
 \be
({\cal F}_{ij} , \; \overline {\cal F}^{ ij})= ({\cal X}_{IJ} , \; \overline {\cal X}^{ IJ}) {\cal V} \ ,
\label{BrinkHoweExact}\ee
since there is only one scalar dependent object,  vielbein,  which transforms as ${\cal V}' =  E^{-1} {\cal V} U(x) $ and makes a bridge between the local $SU(8)$ and global $E_{7(7)}$. The $E_{7(7)}$ symmetry acting on scalars is not deformed, whereas the $E_{7(7)}$ symmetry of the deformed $E_{7(7)}$ must be
$
({\cal X}_{IJ} , \; \overline {\cal X}^{ IJ})' = ({\cal X}_{IJ}, \; \overline {\cal X}^{IJ})  E
$.
 A simple analogy is the relation between a  tangent space vector $V^a=  V^\mu e_{\mu}{}^a$ and a curved space vector $V^\mu$: $V^a$ is invariant under general coordinate transformations but transforms under Lorentz ones. $V^\mu$ is invariant under Lorentz  transformations but transforms under  general coordinate transformations ones. The vielbein $e_{\mu}{}^a$  bridges a linear relation between $V^a$ and $V^\mu$.

  The unique $SU(8)$ and Lorentz covariant, and $E_{7(7)}$ invariant Cremmer-Julia
 constraint
 which reduces the number of real vectors to 28    is
\be\boxed{
{\cal F} _{ij \, \mu\nu} +i \,  {\cal F} _{ij\,  \mu\nu}^* =0.}
\label{CJ}\ee
In this form it  corresponds to eq. (16) in
\cite{Brink:1979nt} and it can  also be presented in  spinor notation as
\be
{\cal F}_{ij \, \mu\nu}= {1\over 2} (\sigma_{\mu\nu})^{\alpha \beta}{\cal M}_{\alpha \beta ij} \, , \qquad \overline {\cal F}^{ij }_{ \mu\nu}= {1\over 2} (\bar \sigma_{\mu\nu})^{\dot \alpha \dot \beta}\overline  {\cal M}_{\dot \alpha \dot\beta}^{ ij} \ .
\ee
Thus, the field strength  ${\cal F}_{ij} $ transforms as a $\underline{\mathbf{28 }}$ under $SU(8)$ and is space-time complex self-dual. The complex conjugate field strength  $\overline {\cal F}^{ij} $ transforms as a ${\mathbf{28 }}$ under $SU(8)$ and is space-time complex anti-self-dual. One can present it as follows
\bea\label{con}
& {\cal F}_{ij \, \alpha  \beta}= {\cal M}_{\alpha \beta ij}\, ,  \qquad &{\cal F}_{ij \, \dot \alpha \dot \beta}=0\, , \nonumber\\
&\overline {\cal F}^{ij }_{ \alpha \beta}=0\, ,  \qquad &\overline {\cal F}^{ij}_{ \dot \alpha \dot \beta}= \overline {\cal M}_{\dot \alpha \dot\beta}^{ ij}\, ,
\eea
and
$
({\cal M}_{\alpha \beta ij})^\dagger = \overline {\cal M}_{\dot \alpha \dot\beta}^{ ij}$, $ (\overline {\cal F}^{ij }_{ \alpha \beta})^\dagger = {\cal F}_{ij \, \dot \alpha \dot \beta}=0
$.
To see how the $SU(8)$ covariant and $E_{7(7)}$ invariant  constraint allows to express $G_{\mu\nu\, IJ}$ as a functional of $F_{\mu\nu}^{IJ}$ and vice versa we need to use the explicit relation between the $E_{7(7)}$ doublets and the $SU(8)$ tensors. From eq. (\ref{BrinkHoweExact})  and (\ref{con}) it follows that
$
{\cal F} _{  ij \dot \alpha \dot \beta} = [(G_{ \dot \alpha \dot \beta} +iF_{ \dot \alpha \dot \beta}) U + (G_{ \dot \alpha \dot \beta} - iF_{ \dot \alpha \dot \beta}) V]_{ij} =0
$
and therefore, as before, in absence of fermions
\be
[ G_{ \dot \alpha \dot \beta} (U+V) +i \, F_{ \dot \alpha \dot \beta}( U-V)] _{ij}  =0\, .
\ee
The constraint (\ref{CJ}) has a unique solution for $G$ in terms of $F$, and vice versa,
since the $U+V$ and $U-V$ scalar-dependent matrices are invertible
\be
G_{ \dot \alpha \dot \beta}= - i F_{ \dot \alpha \dot \beta}  (U-V)( U+V)^{-1} \, , \qquad F_{ \dot \alpha \dot \beta}=  i \, G_{ \dot \alpha \dot \beta}  (U+V)( U-V)^{-1} \ .
\label{constr}\ee
Thus $G$ is a linear function of $F$:  it follows from the unique $SU(8)$ covariant and $E_{7(7)}$ invariant constraint (\ref{CJ}) on 56 $SU(8)$ field strengths. The linear nature of this relation originates in the linear relation between tangent space $SU(8)$ vectors and curved space $E_{7(7)}$  vectors: they are bridged by the vielbein ${\cal V}$ in (\ref{BrinkHoweExact}) and no other relation is possible.

Counterterms independently of details, would violate this requirement, since higher powers of $F$ will be present in  $G$, so $E_{7(7)}$  current conservation forbids them.

 \section{Discussion}
  \setcounter{equation}{0}

 In conclusion, using either the Noether-Gaillard-Zumino $E_{7(7)}$ current conservation or the uniqueness of the Cremmer-Julia constraint we argued that all candidate counterterms of ${\cal{N}}=8$ supergravity are  forbidden. One may wonder whether
it is possible to deform the counterterms  in (\ref{CT})  so that the deformed duality symmetry is respected. The answer is negative: all non-linear counterterms described in
\cite{Howe:1980th,Kallosh:1980fi} provide the unbroken general covariance, local Lorentz and local $SU(8)$ symmetry as well as local supersymmetry.  All this is guaranteed by the fact that in classical theory there is an on shell super-geometry and all torsion and curvature forms satisfy the superspace Bianchi identities  \cite{Brink:1979nt}. Therefore to preserve the undeformed gauge symmetry of the counterterms  we have to use the candidate counterterms constructed in \cite{Howe:1980th,Kallosh:1980fi}. These counterterms are invariant under the undeformed $E_{7(7)}$ symmetry.  However, they are  in conflict with  $E_{7(7)}$ current conservation and the duality symmetry of the deformed equations of motion.

The deep reason why it was possible to construct an infinite number of candidate counterterms  in \cite{Howe:1980th,Kallosh:1980fi} is the fact that the  supersymmetric ``tensor calculus''  for gauge symmetries of the theory is available  \cite{Brink:1979nt}. It is the existence of an ${\cal{N}}=8$ supersymmetric analog (\ref{CT}) of the pure gravity case where there is  a tangent space Riemann-Christoffel curvature tensor $R_{abcd}$ which allows to construct any higher derivative scalars by contracting any number of  curvature tensors with any number of Lorentz covariant derivatives $D_a$ using Minkowski  metric $\eta^{ab}$ so that the number of counterterms proliferates with increasing loop order.

In this paper we have studied   hidden $E_{7(7)}$ duality symmetry between  28-component Bianchi identity,  $\partial_\mu \tilde F^{\mu\nu IJ}=0$, and 28-component
equations of motion of the theory, $\partial_\mu \tilde G^{\mu\nu}_{ IJ}=0 $. Here
$ {\tilde G}_{ IJ}= 2 {\delta (S_0+S_{\rm CT})\over F^{IJ}}=  {\tilde G}_{0 IJ}+  \hat {\tilde G}_{ IJ} $, therefore equations of motion as well as $E_{7(7)}$ duality are deformed by counterterms.
The deformed
$E_{7(7)}$ duality is
\be
\Delta \, \partial_\mu \tilde F^{\mu\nu IJ}= A \, \partial_\mu \tilde F^{\mu\nu IJ}+ B\,  \partial_\mu \tilde G^{\mu\nu IJ} \ .
\ee
Here $A=\mbox{Re}\Lambda-\mbox{Re}\Sigma $,  $B= \mbox{Im}\Lambda+\mbox{Im}\Sigma$ are 133 $E_{7(7)}$ symmetry parameters which mix  the Bianchi identity with deformed equations of motion.
The consistency of the deformed duality
requires an extra infinite number of cancellations for the current conservation in the form of an identity (\ref{critical}), imposed on the deformation of the dual field strength by the  counterterms (\ref{CT}). The identity
 requires that
\be
 {\delta \over \delta F^{IJ} }\int  \hat {\tilde G} B \hat G =0 \ .
\label{critical1}\ee
 Here $ \hat G$ is the deformation of the dual field strength caused by counterterms (\ref{CT}) where
$\hat {\tilde G}^{\mu\nu IJ}= 2 {\delta S_{\rm CT}\over F_{\mu\nu}^{IJ}}$. There is no reason for this infinite number of cancellations, as shown in the paper. As an example, for the sector  $R^2\partial^2 R^2 \partial^2 F $ of the identity (\ref{critical}) for  the 3-loop counterterm the details are given in eq. (\ref{proof}). Based on the analysis of the $E_{7(7)}$ current conservation in the form (\ref{critical1}) we conclude that
 the hidden $E_{7(7)}$ invalidates all gauge symmetry invariant candidate counterterms.

We also presented a unitarity based argument, independent of specific form of $E_{7(7)}$ current conservation  (\ref{critical1}). It requires to use a simple property of all counterterms: they have terms which are at least quartic in $F$.

This argument relies on the uniqueness of the constraint which allows to express the 28-component $E_{7(7)}$ complex doublet $({\cal X}_{IJ} , \; \overline {\cal X}^{IJ})$  via 28 independent real vector fields of the theory. The $SU(8)$ doublet $({\cal F}_{ij} , \; \overline {\cal F}^{ij})$ and the $E_{7(7)}$ doublet  have a unique relation  via the 133-component vielbein ${\cal V}$,  which is
\be
({\cal F}_{ij} , \; \overline {\cal F}^{ ij})= ({\cal X}_{IJ} , \; \overline {\cal X}^{ IJ}) {\cal V} \ .
\label{geom}\ee
The unitarity constraint
 is a unique $E_{7(7)}$ invariant, Lorentz and $SU(8)$ covariant constraint and it can be presented in the form
${\cal F}_{ij \, \alpha  \beta}= {\cal M}_{\alpha \beta ij}\, ,   \overline {\cal F}^{ij}_{ \dot \alpha \dot \beta}=  \overline {\cal M}_{\dot \alpha \dot\beta}^{ ij}({\cal M}_{\alpha \beta ij})^\dagger\, ,
 {\cal F}_{ij \, \dot \alpha \dot \beta}= \overline {\cal F}^{ij }_{ \alpha \beta}=0$, see the derivation in (\ref{con}).
This constraint is not valid when counterterms with higher   powers of $F$ are added to the action, since it leads to a non-linear relation between $SU(8)$ doublet  and $E_{7(7)}$ complex doublet, in contradiction with the ${E_{7(7)}\over SU(8)}$ coset space geometry relation (\ref{geom}). The details can be found in Sec. 5.

Thus, in the absence of anomalies perturbative ${\cal{N}}=8 $ supergravity is predicted to be UV finite, in agreement with the light-cone superspace
prediction \cite{Kallosh:2010kk}.


\section*{Acknowledgments}
We are grateful to  J. Broedel, D. Freedman, S. Kachru, A. Linde,  E. Silverstein, L. Susskind, A. Van Proeyen  and H. Verlinde for stimulating  discussions.
This work  is supported by the NSF grant 0756174.

\appendix
\section{Derivation of the NGZ identity  \cite{Gaillard:1981rj,Aschieri:2008ns}}
 \setcounter{equation}{0}
Note that
\be
S[F', \varphi]- S[F, \varphi]=\int  \Delta F {\delta S\over \delta F}= {1\over 2} \int \Delta F \tilde G \ .
\ee
It is also equal to $S[F', \varphi']- S[F, \varphi']$ since we are making infinitesimal transformations.
It follows that
\be
{\delta \over \delta F^\Lambda }\Big (  S[F', \phi']- S[F, \varphi'] \Big )= {1\over 2}{\delta   \Delta F \over \delta F^\Lambda } \tilde G +   {1\over 2}\Delta F {\delta  \tilde G  \over \delta F^\Lambda } \ .
\label{2}\ee
We  also compute
\be
{\delta \over \delta F^\Lambda }\Big ( S[F, \varphi']- S[F, \varphi]\Big )= \Delta \varphi  {\delta^2 S\over \delta F^\Lambda \delta \varphi }= {1\over 2} \Delta \varphi  {\delta \tilde G \over  \delta \varphi } \ .
\label{3}\ee
Now we sum (\ref{2}) and (\ref{3}) and we get
\be
{\delta \over \delta F^\Lambda }\Big ( S[F', \varphi']- S[F, \varphi]\Big )=  {1\over 2} \Delta \varphi  {\delta \tilde G \over  \delta \varphi }+   {1\over 2}\Delta F {\delta  \tilde G  \over \delta F^\Lambda }  +  {1\over 2}{\delta   \Delta F \over \delta F^\Lambda } \tilde G \ .
\label{4}\ee
Now we use the following:
\be
 \Delta \varphi  {\delta \tilde G \over  \delta \varphi }+   \Delta F {\delta  \tilde G  \over \delta F^\Lambda }
  =   \Delta \tilde G = C\tilde F + D\tilde G
  \label{5}\ee
since
\begin{eqnarray}\label{tilde}
\Delta \left(
                      \begin{array}{cc}
                 \tilde  F \\
                     \tilde G \\
                      \end{array}
                    \right)\ =\left(
                      \begin{array}{cc}
                      A&  B \\
                        C & D \\
                      \end{array}
                    \right)  \left(
                      \begin{array}{cc}
                     \tilde F \\
                       \tilde G \\
                      \end{array}
                    \right) \ .
\end{eqnarray}
We may now continue with eq. (\ref{4})
\be
{\delta \over \delta F^\Lambda }\Big ( S[F', \varphi']- S[F, \varphi]\Big )=  {1\over 2} (C\tilde F + D\tilde G)  +  {1\over 2}( A ^T \tilde G +  {\delta G\over \delta F} B \tilde G)= {1\over 2} (C\tilde F  +    {\delta G\over \delta F} B \tilde G) \ .
\label{5}\ee
Here we took into account the properties of the $Sp(2n,{\mathbb{R}})$ transformations
\begin{eqnarray}
 A^{\mathsf{T}} =-D \ ,   \qquad B^{\mathsf{T}}= B \ ,  \qquad C^{\mathsf{T}}=C \ .
\end{eqnarray}
Note that
\be
 {1\over 2}(C\tilde F  +    {\delta G\over \delta F}  B \tilde G)= {1\over 4} {\delta \over \delta F^\Lambda }
 (F C\tilde F  +   G B   \tilde G) \ ,
\ee
which proves that
\be
{\delta \over \delta F^\Lambda }\Big ( S[F', \varphi']- S[F, \varphi]-  {1\over 4} (F C\tilde F  +   G B   \tilde G) \Big )=0 \ .
\label{6}\ee
{\it NGZ conserved current}

 In the $SU(8)$ local version of the theory the Noether current of scalars has an elegant form which follows from the action (in absence of fermions)
\begin{eqnarray}\label{lag2}
L_{\cal V}=-\frac{1}{2}\mbox{Tr}\Big((D_{\mu}{\cal V}){\cal V}^{-1} (D^{\mu}{\cal V}){\cal V}^{-1}\Big) \ .
\end{eqnarray}
\be
j_{\cal V}^\mu= \rm  Tr \left ( {\partial L\over \partial   {\cal V}_{, \mu}} {\cal V}\, \delta E^{-1}\right )  = -  \rm  Tr \Big({\cal V}^{-1} (D^\mu {\cal V})  \,\delta E^{-1}\Big ) \ .
\ee
We may now introduce the Gaillard-Zumino current \cite{Gaillard:1981rj}
\be
\hat J^\mu \equiv  {1\over 2} \left (\tilde G^{\mu\nu} A \, {\cal A}_\nu -\tilde F^{\mu\nu} C {\cal A}_\nu +\tilde G^{\mu\nu}  B {\cal B}_\nu -\tilde F^{\mu\nu} D {\cal B}_\nu  \right) \ ,
\ee
whose divergence cancels the scalar variation of the Lagrangian when equations of motion are satisfied. The  classical Lagrangian provides the conservation of the total current, the Noether current of the scalars and the Gaillard-Zumino current of vectors:
$
\partial_\mu J^\mu= \partial_\mu \hat J^\mu+ \partial_\mu j^\mu_{\cal V}=0$.

\section{Counterterms and the Hamiltonian approach to duality}
 \setcounter{equation}{0}

In the Hamiltonian approach\footnote{There is a related  issue in   \cite{Bossard:2010dq} where the $E_{7(7)}$ symmetry is realized off shell in the Lorentz non-covariant way.} to duality symmetries  \cite{Bunster:2011aw}, for example in the Coulomb gauge, it is important  that in the classical action $A_0$ is a Lagrange multiplier \be
L= \pi^i \dot A_i- H(\pi^i, A_i) + A_0 \, \partial_i \pi^i \ ,
\ee
and therefore the momenta $\pi^i$ conjugate to the vector $A_i$ satisfies the constraint $\partial_i \pi^i=0$.
The resolution of this constraint requires to introduce in addition to $A_i$ the second vector potential $Z_i$
\be
\pi^i= -{1\over 2} \epsilon^{ijk} (\partial _j Z_k- \partial _k Z_j) \ .
\ee
The fact that the Hamiltonian constraint  $\partial_i \pi^i=0$ is the Gauss' law  is fundamental in duality symmetric theories. It explains the doublet nature of  potentials in ungauged supergravity theories $(A_i,\,  Z_i)$, where scalars and fermions interact with vectors only via $F_{\mu\nu}$.

The counterterms in ${\cal{N}}=8$ supergravity  \cite{Howe:1980th,Kallosh:1980fi} depend on   $ F_{\mu\nu}$,  $D_\lambda F_{\mu\nu}$,  $D_\delta D_\lambda F_{\mu\nu}$ etc.
If any of such counterterms were added to the classical action,  $A_0$ would not  be a Lagrange multiplier anymore.  It is not clear {\it a priori}  if the Gauss law and second vector potential required for duality are available.  In presence of  deformation of the classical action by candidate counterterms  with derivatives of $F_{\mu\nu}$ the analysis of dualities in \cite{Bossard:2010dq}, \cite{Bunster:2011aw} needs to be revisited to find out how the deformations affect duality symmetry of the Hamiltonian/Lorentz non-covariant action. Such analysis will lead to an independent statement about
 the implications of the continuous global non-compact $E_{7(7)}$ duality on perturbative ${\cal{N}}=8$ supergravity.


\end{document}